\documentclass[aps, prd, twocolumn, showpacs, superscriptaddress, groupedaddress]{revtex4-1} 
\usepackage{graphicx}    
\usepackage{amssymb}
\usepackage{dcolumn}
\usepackage{color}
\usepackage{subfigure, rotating, bm, array}
\usepackage[pagebackref=false, colorlinks=true]{hyperref}
\hypersetup{
linkcolor=blue,     % color of internal links
citecolor=blue,     % color of links to bibliography
urlcolor=blue}   
\usepackage{amsmath}
\usepackage{ulem}
\usepackage[utf8]{inputenc}

\begin{document}
\title{Singularity resolution in gravitational collapse}

\author{Karim Mosani}
\email{kmosani2014@gmail.com}
\affiliation{BITS Pilani K.K. Birla Goa Campus, Sancoale, Goa 403726, India}

\author{Dipanjan Dey}
\email{dipanjandey.icc@charusat.ac.in}
\affiliation{International Center for Cosmology, Charusat University, Anand 388421, Gujarat, India}

\author{Kaushik Bhattacharya}
\email{kaushikb@iitk.ac.in}
\affiliation{Department of Physics, Indian Institute of Technology, Kanpur, Kanpur 208016, India}

\author{Pankaj S. Joshi}
\email{psjprovost@charusat.ac.in}
\affiliation{International Center for Cosmology, Charusat University, Anand 388421, Gujarat, India }

\date{\today}
\begin{abstract}
We investigate the unhindered gravitational collapse of a homogeneous scalar field with nonzero potential, a two-dimensional analog of the Mexican hat-shaped Higgs field potential. The collapsing scalar field is surrounded by an exterior retarded (outgoing) generalized Vaidya spacetime. We prove that the density dependence on the scale factor cannot be expressed as an algebraic function in such a scenario. For a certain transcendental expression of the density of such field as a function of scale factor, we then show that the collapse evolves to a singularity at an infinite comoving time, which is equivalent to saying that the singularity is avoided altogether. An ultra high density region of the order of Planck length can, however, be reached in a finite comoving time. The absence of the formation of trapped surfaces makes this ultra high density region globally visible.\\

\textbf{key words}: Gravitational Collapse, Naked Singularity, Scalar field

\end{abstract}
\maketitle
\section{Introduction}
Hawking and Penrose depicted the formation of singularities in the gravitational collapse and cosmology in what is known as the singularity theorems
\cite{Hawking_73, Wald_84}. 
However, it is widely believed that no past inextendible nonspacelike geodesic can exist between the singularity and any point on the spacetime manifold.
%The singularity formed due to gravitational collapse is preferred not to have a causal connection from `it' to any point on the spacetime manifold. 
In other words, no nonspacelike geodesics could have a positive tangent at the singularity. This statement is known as the strong cosmic censorship hypothesis
\cite{Penrose_69, Joshi_07}. One motivation tempting us towards the strong cosmic censorship hypothesis is our desire always to have a globally hyperbolic spacetime metric
\cite{Joshi_07}. 

Global hyperbolicity is the most potent form of the causality condition
\cite{Wald_84}. 
Even if the spacetime is strongly causal, which ensures no formation of closed nonspacelike curves, such a spacetime metric should retain the property of no violation of causality if there is a small perturbation in the spacetime metric. Thus, there should be stability in the spacetime metric in the sense that all ``nearby” spacetime metrics should also have a similar property as far as the causality condition is concerned. However, different topologies can be defined on the set of all Lorentz metrics, each giving a different meaning to the word ``nearby.” This drawback can be overcome by considering the spacetime to be globally hyperbolic, an alternative, and the strongest description of the causality condition. Globally hyperbolic spacetime harbors a unique topology that is homeomorphic to $ S\times \mathbb{R}$, where $S$ is a Cauchy surface admitted by the spacetime manifold
\cite{Geroch_70}
(any two Cauchy surfaces are homeomorphic to each other). A global hyperbolic spacetime is stably causal
\cite{Hawking_73}. 
One can show that if the strong cosmic censorship is not valid, then the spacetime metric is not globally hyperbolic
\cite{Joshi_82}.   
Nevertheless, one can also show that a sufficiently inhomogeneous spherically symmetric dust cloud collapse can end up in a singularity such that null geodesics can escape from the singularity without getting trapped by the trapped surfaces
\cite{Joshi_93, Mosani_20(2)}. 
Such singularities are also stable under small perturbation in a certain subset of the entire initial data leading to the collapse
\cite{Deshingkar_98, Mena_2000, Mosani_20}. 
These may act as a counterexample to the strong cosmic censorship hypothesis. 
Apart from the scenario mentioned above, examples of gravitational collapse of various matter fields have been shown to give rise to a naked singularity
\cite{Christodoulou_91, Magli_97, Magli_98, Harada_99, Harada_02, Goswami_04, Goswami_04b, Giambo_03, Giambo_06}. 
The intriguing question is whether cosmic censorship is respected in the gravitational collapse of fundamental matter fields deduced from a suitable Lagrangian. 

In the case of a massless scalar field, it was shown by Christodoulou that the cardinality of the linearly independent elements in the set of initial data giving rise to a naked singularity as an end state is strictly less than that of the entire collection of initial data
\cite{Christodoulou_94, Christodoulou_99}. 
In other words, in the case of scalar field collapse with zero $\phi^2$ term in the scalar field Lagrangian, the set of initial data giving rise to naked singularity has positive codimension in the entire set of initial data (the whole set includes those initial data giving rise to a black hole, and those giving rise to a naked singularity). This outcome concludes that a massless scalar field collapses to a naked singularity that is nongeneric (by the generic outcome of the gravitational collapse, we mean that the set of the initial data giving rise to the outcome has a nonzero measure in the entire initial data set
\cite{Joshi_07}). 
Genericity aspects of the naked singularity formed due to massive scalar field collapse are yet to be studied. 

In \cite{Goswami_04b}, a massive homogeneous scalar field having a certain potential $V(\phi) \propto e^{-\phi}$ is shown to form a naked singularity at the end of its collapse. Here, we are interested in a more realistic potential of the scalar field, whose Lagrangian has a $\mathbb{Z}_2$ symmetry. Symmetries in scalar field theories are interesting as we know the Higgs field potential \cite{Higgs_64} also has symmetries. In particle physics, the Higgs field plays a pivotal role in determining the properties of the matter content of the known universe. In our case, we work with a much-simplified model consisting of only one real scalar field whose potential has a particular symmetry. Such potential is motivated by the theory of phase transitions by Landau \cite{Melo_17}. At temperatures above a certain cut-off $T_c$, the scalar field has zero average value in its lowest energy state, i.e., the vacuum.  At temperatures below $T_c$, such a field has two nonzero average values in vacuum (the value of these two states are the same in magnitude but differ in polarity). As there appears multiple vacua in the theory below a certain temperature, the system has to choose any one of the vacua, thereby breaking the symmetry of the system. All the perturbations of the field now have to be done with respect to a particular vacuum, and consequently, the overall symmetry is broken by the vacuum. This mode of symmetry breaking is known as spontaneous symmetry breaking.  In particle physics, a considerable amount of vacuum energy in the universe is contributed by the Higgs field due to its nonzero average lowest energy state \cite{Rugh_02}. In the forthcoming sections, we will discuss the dynamics of gravitational collapse in the presence of a scalar field which shows spontaneous symmetry breaking.

In this paper, we show that for a certain homogeneous density dynamics (gravitational collapse) of our toy model scalar field with a potential which has $\mathbb{Z}_2$ symmetry, the cloud collapses eternally, thereby the singularity formation is avoided. However, a very ultra high density region (UHDR) of the size of the order of Planck length is obtained in a finite comoving time. Trapped surfaces do not form, causing the exposure of this strong gravity region to the outside observer. 

The paper is arranged as follows: in Sec. II, we derive basic dynamical equations of the massive scalar field using the Einsteins field equations, which will be used in further investigations. In Sec. III, we show that for the scalar field mentioned above, the density cannot be an algebraic function of the scale factor but a transcendental function. For a particular transcendental function as a density configuration, satisfying the regularity requirements, we obtain the exact solution of Einstein's field equations. The exterior spacetime is modeled by the retarded (outgoing) generalized Vaidya spacetime. We depict the conditions that need to be satisfied for smoothly matching the interior collapsing solution with the exterior generalized Vaidya metric. In Sec. IV we discuss the visibility aspects of the UHDR, which may be governed by quantum gravity, and see if $\mathbb{Z}_2$ symmetry in the scalar field potential is obeyed by the causal property of the UHDR. We end the paper by concluding remarks and final discussions.  We use the geometrized units $c=8\pi G=1$.

\section{Einstein's field equations and collapse dynamics}
Consider a homogeneous gravitational collapse of the perfect fluid scalar field $\phi=\phi(t)$ having the potential $V(\phi)$. The components of the stress-energy tensor are given by
\begin{equation}
    T^{\mu}_{\nu}=\textrm{diag}\left(\rho,p,p,p\right).
\end{equation}
 The spacetime geometry is governed by the Friedmann–Lemaître–Robertson–Walker (FLRW) metric
\begin{equation}
  ds^2=-dt^2+ a^2dr^2+R^2d\Omega^2\,,
\end{equation}
where $d\Omega^2=d\theta^2 + \sin^2\theta d\phi^2$. Here $a=a(t)$ is the scale factor such that $a(0)=1$ and $a(t_s)=0$, where $t_s$ is the time of formation of the singularity. $R=R(t,r)$ is the physical radius of the collapsing cloud and can be written as
\begin{equation}
    R(t,r)=r a(t).
\end{equation}
The Lagrangian of the scalar field is given by
\begin{equation}
    \mathcal{L}_{\phi}=\frac{1}{2}g^{\mu \nu}\partial_{\mu}\phi \partial_{\nu} \phi-V(\phi),
\end{equation}
The stress-energy tensor is then 
\begin{equation}
    T_{\mu\nu}=-\frac{2}{\sqrt{-g}}\frac{\delta \left(\sqrt{-g} \mathcal{L}_{\phi}\right)}{\delta g^{\mu\nu}}.
\end{equation}
The density and the isotropic pressure are subsequently expressed in terms of the time derivative of the scalar field and its potential as
\begin{equation}\label{rho}
    \rho= \frac{1}{2}\dot \phi^2+V(\phi)=\frac{3\dot a^2}{a^2}, 
\end{equation}
and
\begin{equation}\label{p}
    p=\frac{1}{2}\dot \phi^2-V(\phi)=-\frac{2\ddot a}{a}-\frac{\dot a^2}{a^2}.
\end{equation}
The overhead dot denotes the time derivative of the functions. The Klein-Gordan equation
\begin{equation}\label{KG}
    \ddot \phi+\frac{3\dot a}{a} \dot \phi+V_{,\phi}=0,
\end{equation}
can be obtained from the Einstein's field Eqs.(\ref{rho}) and (\ref{p}) (or from Eq.(\ref{rho}) along with the Bianchi identity) if $\dot \phi$ does not vanish identically. Hence the Klein-Gordan equation should not be seen as an independent equation constraining the choice of free functions. From Eqs.(\ref{rho}) and (\ref{p}), and from using the chain rule $\dot  \phi=\phi_{,a} \dot a$, we get

\begin{equation}\label{rhop}
    \rho+p=\phi_{,a}^2\dot a^2.
\end{equation}
Equation (\ref{rho}) can be rewritten to obtain the dynamics of the collapse as
\begin{equation}\label{adot}
    \dot a=-\sqrt{\frac{\rho(a)}{3}}a,
\end{equation}
differentiating which, we obtain
\begin{equation}\label{addot}
    \ddot a=\frac{1}{3}a\left(\frac{a\rho_{,a}}{2}+\rho \right).
\end{equation}
Using Eq.(\ref{adot}) in Eq.(\ref{rhop}), we get
\begin{equation}\label{rhop2}
    \rho\left(1-\frac{\phi_{,a}^2a^2}{3}\right)+p=0.
\end{equation}
From Eqs.(\ref{rho}) and (\ref{adot}), we get
\begin{equation}\label{rhop3}
    p=\rho-2V.
\end{equation}
Using Eqs.(\ref{rhop2}) and (\ref{rhop3}), we get
\begin{equation}\label{rhofinal}
    \rho=\frac{V(\phi)}{1-\frac{\phi_{,a}^2a^2}{6}}.
\end{equation}
Using Eqs.(\ref{p}), (\ref{adot}), and (\ref{addot}) in Eq.(\ref{rhop}) and rearranging, we obtain
\begin{equation}\label{rhoa}
    \frac{\rho_{,a}}{\rho}=-\phi_{,a}^2 a.
\end{equation}
Differentiating Eq.(\ref{rhofinal}) with respect to $a$ and substituting in Eq.(\ref{rhoa}), we obtain a second order non-linear differential equation 
\begin{equation}
    \frac{V_{,\phi}\phi_{,a}}{V}+\frac{a}{3}\frac{\phi_{,a}\left(a\phi_{,aa}+\phi_{,a}\right)}{1-\frac{\phi_{,a}^2a^2}{6}}+a \phi_{,a}^2=0.
\end{equation}
Now, in case $\phi_{,a}\neq0$, we can get the reduced form of the above differential equation as
\begin{equation}\label{dephi}
    \frac{V_{,\phi}}{V}+\frac{a}{3}\frac{\left(a\phi_{,aa}+\phi_{,a}\right)}{\left(1-\frac{\phi_{,a}^2a^2}{6}\right)}+a \phi_{,a}=0.
\end{equation}
For a given potential $V=V(\phi)$, one can solve the differential Eq.(\ref{dephi}) by choosing two suitable initial conditions to get $\phi=\phi(a)$. In the next section, we derive the exact solution of the collapsing toy model scalar field with potential having $\mathbb{Z}_2$ symmetry.
\begin{figure*}\label{fig1}
\subfigure[]
{\includegraphics[scale=0.5]{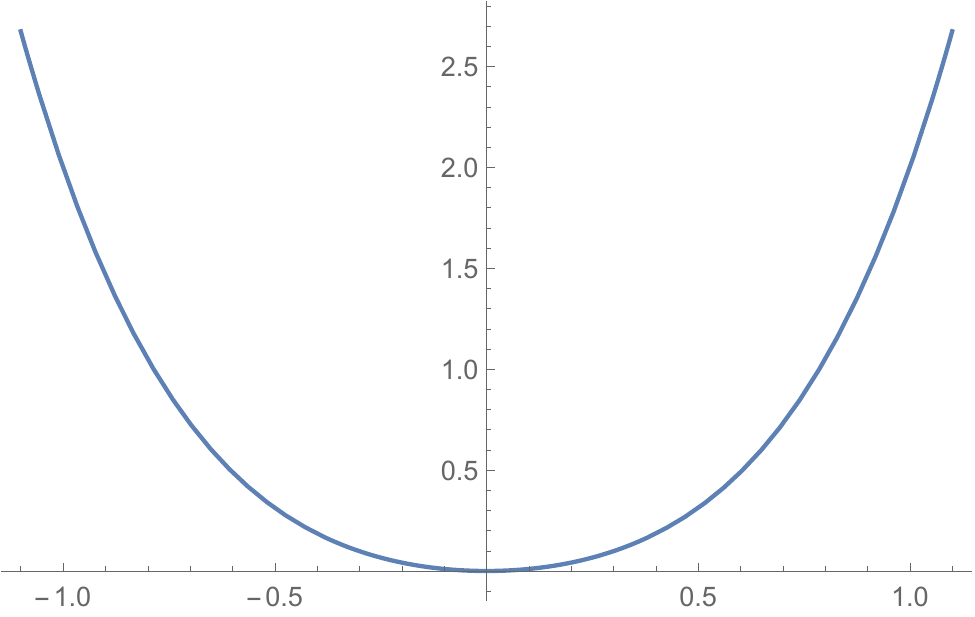}}
\hspace{0.2cm}
\subfigure[]
{\includegraphics[scale=0.5]{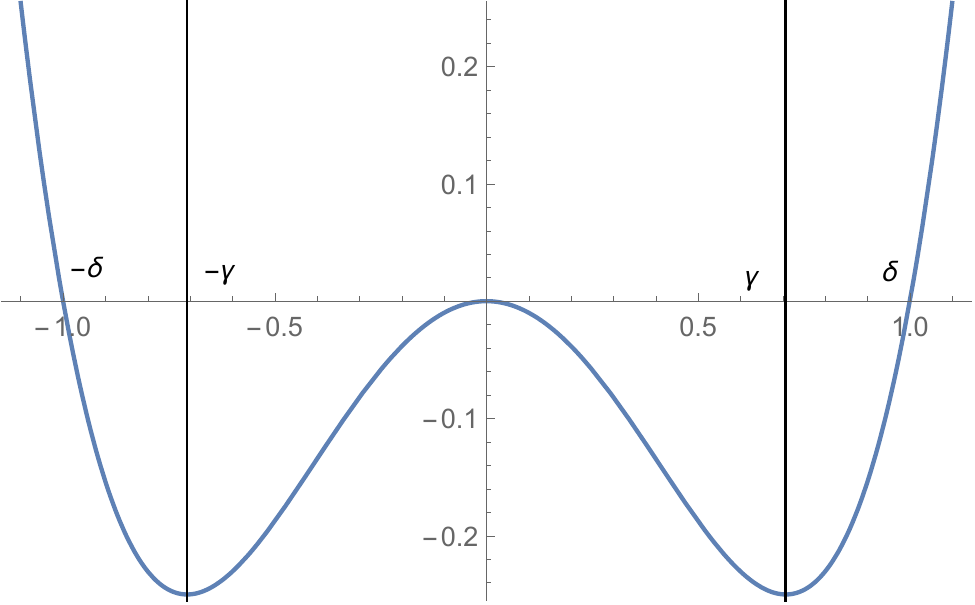}}
\caption{The scalar field $\phi$ and the potential $V(\phi)$ are represented by the horizontal and vertical axis respectively. $q>0$ in $(a)$ and $q<0$ in $(b)$. Here, $\gamma=\sqrt{\frac{-q}{4\lambda}}$ and $\delta=\sqrt{\frac{-q}{2\lambda}}$.}
\end{figure*}
\section{Exact solution: Eternally collapsing cloud}
%\textbf{Theorem}: Singularity formed due to unhindered gravitational collapse of homogeneous massive scalar field with potential in Eq.(\ref{V}) is always completely hidden.
The toy model scalar field which we consider, is described by the potential (Fig.(1))
 \begin{equation}\label{V}
     V(\phi)=\frac{1}{2} q \phi^2+\lambda \phi^4.
\end{equation}
%the scalar field Lagrangian 
%\begin{equation}
%    \mathcal{L}=\frac{1}{2}g^{\mu \nu}\partial_{\mu}\phi \partial_{\nu} \phi-V(\phi)
%\end{equation}
%is invariant under the reflection symmetry transformation $\phi \to -\phi$. Depending on the polarity of $q$, we have different reflection symmetries of $V(\phi)$ as shown in Fig.(1).
At high temperature $T$, above a certain cutoff $T_c$, $q>0$. The potential in such case is a parabola (see Fig.(1.a)), with the average of its lowest energy state at 
\begin{equation}
    \phi_v=0.
\end{equation}
This is the vacuum. As the temperature cools down below $T_c$, $q<0$, and the potential is no more a parabola (see Fig.(1.b)). Now there are two states of vacuum, i.e. the lowest energy states, represented by  
\begin{equation}\label{nonzerovacuum}
    \phi_v=\pm \sqrt{-\frac{q}{4\lambda}}.
\end{equation}
Both the vacuum states are allowed to exist with equal probability. One can see that at $T>T_c$, the lowest energy state is symmetric with respect to $\phi \longleftrightarrow -\phi$ reflection. However, at $T<T_c$, choosing any one of the ground states as the lowest energy state, we see that the symmetry  $\phi \longleftrightarrow -\phi$ is not respected any more.

Now, as far as solving the Einstein's field equations are considered, we can, in principle, get the complete solution by solving the differential Eq.(\ref{dephi}) for the particular potential (\ref{V}). This will give us the functional form of the scalar field $\phi$ in terms of the scale factor. Once $\phi(a)$ is obtained, we can get $\rho(a)$ from Eq.(\ref{rhofinal}), $p(a)$ from Eq.(\ref{rhop3}), and $a(t)$ from Eq.(\ref{adot}), thereby, solving the field equations completely.
%We can get a differential equation by using Eq.(\ref{V}), and its derivative with respect to $\phi$, and substituting in the differential Eq.(\ref{dephi}). Then, we can, in principle, get the solution, an expression of $\phi$ in terms of $a$, which corresponds to the potential in Eq.(\ref{V}). 
However, the problem with this particular approach is that the differential Eq.(\ref{dephi}) thus obtained, is non-linear and is not easy to solve analytically. We, therefore, don't proceed with this approach. Instead, we carefully choose the density configuration as a function of the scale factor such that the underlying potential turns out to be of the form in Eq.(\ref{V}). This can be done due to one degree of freedom available to us. 

However, choosing a particular density configuration so as to ensure that the scalar field follows the ``Higgs like" potential (Eq.(\ref{V})) is a hit and trial method. It is obviously not practical to try each and every possible kind of density configuration and see if the corresponding underlying potential of the scalar field is the one that we desire. The following theorem, however,  helps us to choose a suitable density profile that can lead to a desirable potential. It states: In a gravitational collapse involving homogeneous scalar field $\phi(a)$ governed by FLRW spacetime, if $\rho(a)$ is an algebraic function, then either $V(\phi)$ is a transcendental function, or  
    \begin{equation}\label{Vorderalpha}
        \lim_{a \to 0}V(\phi(a))=\frac{1}{\phi^{\alpha}}; \hspace{0.2cm} \alpha>0.
    \end{equation}
    (If (\ref{Vorderalpha}) is true, then $V(\phi)$ can be algebraic or transcendental.)
    
%Now consider the potential mentioned in Eq.(\ref{V}). Substituting Eq.(\ref{phiorder2}) in Eq.(\ref{V}), we obtain
%\begin{equation}
%    \lim_{a\to 0} V \sim O\left(a^{n-m}\right).
%\end{equation}
%This also does not match with the order of $V$ we expect, i.e. (\ref{Vorder}), when the density rises as (\ref{rhoorder}) $\hspace{2cm}\Box$.

%Hence we conclude that for a massive scalar field has the potential, as mentioned in Eq.(\ref{V}), the density configuration is not an algebraic function in $a$. It, therefore, has to be transcendental in nature. The result is valid for both cases: $q<0$ and $q>0$. 

The proof of this theorem can be found in the appendix A. One of the results of this theorem is that the density configuration of the collapsing homogeneous scalar field having the potential (\ref{V}) is not an algebraic function of $a$. The proof of this statement can be found in the appendix A. Employing this fact, let us set the density configuration of the homogeneous scalar field as the following transcendental function in $a$: 
\begin{equation}\label{rhoeternal}
    \rho(a)=64\lambda(k-\log a)^2, \hspace{0.2cm} k>0.
\end{equation}
For such density profile, using Eqs.(\ref{adot}, \ref{addot}, \ref{p}) in Eq.(\ref{rhop}), we get 
\begin{equation}\label{phiaeternal}
    \phi_{,a}^2=\frac{2}{a^2\left(k-\log a\right)}.
\end{equation}
Taking the square root gives us
\begin{equation}
     \phi_{,a}=\pm \frac{\sqrt{2}}{a\sqrt{k-\log a}}.
\end{equation}
Integrating, the above equation, we get
\begin{equation}\label{phieternal}
    \phi(a)=\mp 2\sqrt{2}\sqrt{k-\log a}.
\end{equation}
Using Eq.(\ref{rhoeternal}) and Eq.(\ref{phiaeternal}) in Eq.(\ref{rhofinal}), we then obtain
\begin{equation}\label{V2}
    V(\phi)=-\frac{8}{3}\lambda \phi^2+\lambda \phi^4,
\end{equation}
which resembles the potential depicted in Eq.(\ref{V}) with 
\begin{equation}\label{q}
q=\frac{16}{3}\lambda.    
\end{equation}
If we choose $\phi_{,a}>0$, then $\phi<0$, and vice versa. We can see that as the cloud evolves from $a=1$ to $a=0$ (singularity), the scalar field evolves from $-2\sqrt{2}\sqrt{k}$ to $-\infty$ if we choose $\phi_{,a}>0$. It evolves from $2\sqrt{2}\sqrt{k}$ to $\infty$ if we choose $\phi_{,a}<0$. 

As a special case example, for $k=k_v$ given by 
\begin{equation*}
    k_v=\frac{1}{6},
\end{equation*}
the collapse initiates from the vacuum state 
\begin{equation}
\left(\phi_v, V_v\right)=\left(\pm \frac{2}{\sqrt{3}},-\frac{23 \lambda}{9} \right),    
\end{equation}
Eq.(\ref{nonzerovacuum}). It ends at the singularity, for which the scalar field and the potential blows up as
\begin{equation}
    \lim_{a\to 0}\left(\phi, V\right)=\left(\pm \infty, \infty \right).
\end{equation}

We, therefore, have a class of scalar field collapse with the potential Eq.(\ref{V2}), which starts from some finite nonzero value of $\phi$ and blows up in the end, maintaining the polarity of $\phi$ throughout the collapse. It is worth mentioning that Eq.(\ref{phieternal}) and Eq.(\ref{V2}) satisfies the differential Eq.(\ref{dephi}).

To understand the time evolution of the cloud, we solve differential Eq.(\ref{adot}) by substituting Eq.(\ref{rhoeternal}) and setting the constant of integration such that $a(0)=1$. We then obtain 
\begin{equation}\label{scalefactor}
    a(t)=\exp\left(k\left(1-\exp\left(\frac{8\sqrt{\lambda}t}{\sqrt{3}}\right)\right)\right).
\end{equation}
We can see that $a\to 0$ at $t\to \infty$. Hence, the singularity is formed at the infinite comoving time. As mentioned before, the UHDR is, however, reached in a finite comoving time. We will discuss the property of this region in the next section. But before, we would like to note that the eternally collapsing solution derived here, which we call an interior metric, can be joined smoothly with an exterior generalized Vaidya metric
\cite{Glass_98, Wang_99}
at the boundary of the collapsing cloud such that the union of these two metrics forms a valid solution to Einstein's field equations
\cite{footnote1}.
%\footnote{Even if there is a discontinuity or jump in the curvature term at $\Sigma$, which violates the junction conditions
%\cite{Darmios_27, Israel_67}, 
%one could give its physical interpretation in the sense that there exists surface stress-energy term on $\Sigma$. Such a scenario should not be considered unphysical
%\cite{Poisson_04}.}. 
The boundary is a timelike hypersurface $\Sigma$ having the radial coordinate $r_c$.  We rewrite the metric governing the eternally collapsing scalar field (the interior metric) as
\begin{equation}
    ds^{2}_{-}=-dt^2+a^2 dr^2+R^2 d\Omega ^2.
\end{equation}
 The exterior generalized Vaidya spacetime is expressed as
\begin{equation}\label{gvs}
    ds^2_{+}=-\left(1-\frac{2\mathcal{M}(R_c,v)}{R_c}\right)dv^2-2dvdR_c+R^2_c d\Omega^2.
\end{equation}
Here $v$ is the retarded (outgoing) null coordinate, $R_c$ $(=R_c(t))$ is the Vaidya radius, and $\mathcal{M}(R_c,v)$ is the generalized Vaidya mass function. Smooth matching of the interior and exterior spacetime at $\Sigma$ was discussed in detail in 
\cite{Goswami_04b}. 
Following their work, the relevant equations obtained by matching the first and second fundamental forms for the interior and the exterior metrics on $\Sigma$, are as follows:
\begin{equation}\label{mc1}
    R_c(t)=R(t,r_c) \left(=r_c a(t)\right),
\end{equation}
\begin{equation}\label{mc2}
    F(t,r_c)=2\mathcal{M}(R_c,v),
\end{equation}
\begin{equation}\label{mc3}
    \left(\frac{dv}{dt}\right)_{\Sigma}=\frac{1+\dot R_c }{1-\frac{F(t,r_c)}{R_c}},
\end{equation}
and
\begin{equation}\label{mc4}
    \mathcal{M}(R_c,v),_{R_c}=\frac{F(t,r_c)}{2 R_c}+R_c \ddot R_c.
\end{equation}
Here, $F=F(t,r)= R \dot R^2$ is the Misner-Sharp mass function of the collapsing scalar field. Assurance of smooth matching of the two spacetime metrics on $\Sigma$ puts a restriction on the otherwise free Vaidya generalized mass function $\mathcal{M}(R_c,v)$ in the sense that the above four equations should be satisfied on $\Sigma$. Using Eq.(\ref{addot}), Eq.(\ref{rhoa}), Eq.(\ref{rhoeternal}), and Eq.(\ref{phiaeternal}), along with Eq.(\ref{mc1}) and Eq.(\ref{mc2}), we can rewrite Eq.(\ref{mc4}) as
\begin{equation}\label{deM}
    \mathcal{M},_{R_c}-\frac{\mathcal{M}}{R_c}-A R_c^2\left(B-\log R_c \right)\left(B-1-\log R_c \right)=0,
\end{equation}
where
\begin{equation}\label{MAB}
    \mathcal{M}=\mathcal{M}(R_c,v), \hspace{0.2cm} A=\frac{64\lambda}{3}, \hspace{0.2cm} \textrm{and} \hspace{0.2cm} B=k+\log r_c.
\end{equation}
Solving the differential Eq.(\ref{deM}), one can obtain the dependence of $\mathcal{M}$ on $R_c$ as
\begin{equation}\label{gvmf}
    \mathcal{M}=\frac{1}{2}A R_c^3\left(B-\log R_c\right)^2+R_c \mathcal{M}_1(v).
\end{equation}
Here, the function $\mathcal{M}_{1}(v)$ is still free, hence this expression rather describes a class of exterior generalized Vaidya mass function, which allows smooth matching of the exterior generalized Vaidya spacetime with the interior eternally collapsing ball of homogeneous scalar field with potential mentioned in Eq.(\ref{V2}).

Here, one could ask the reason for choosing the generalized Vaidya solution for modeling the exterior spacetime. If the exterior region is governed by the Schwarzschild metric, which is a particular case of Eq.(\ref{gvs}) having constant $\mathcal{M}$, then smooth matching is not possible as is apparent from Eq.(\ref{mc1}-\ref{mc4}). A physical explanation would be to say that the negative pressure inside the collapsing cloud causes diffusion of matter field and the outward flowing photons (called null dust) in the outer region. The internal pressure of the collapsing cloud is related to $F$ as 
\begin{equation} \label{penergitics}
    p=-\frac{\dot F}{R^2 \dot R}.
\end{equation}
Now, using Eq.(\ref{rho}), Eq.(\ref{p}), Eq.(\ref{addot}), and Eq.(\ref{rhoeternal}), one can express the dynamics of pressure as
\begin{equation}\label{pa}
    p=\frac{128}{3} \lambda \left(k-\log{a}\right)-64 \lambda \left(k-\log{a}\right)^2.
\end{equation}
One can observe from the above equation that given a $\lambda$ and $k$, after a certain time through the process of collapse, the pressure becomes negative and remains so after that. The diffusion of matter from the ever-collapsing scalar field is due to this negative pressure inside the cloud, which causes the Misner-Sharp mass function of the cloud to decrease with time, as seen from Eq.(\ref{penergitics}). Hence, there is no vacuum in the exterior of the collapsing cloud.  Therefore, this outer region can be modeled by spacetime corresponding to the stress-energy tensor, the linear superposition of null dust, and perfect fluid. Such spacetime is nothing but the generalized Vaidya spacetime. It is a wider class of spacetime, including many known solutions like the Vaidya solution
\cite{Vaidya_51}, 
monopole solution
\cite{Barriola_89}, 
charged Vaidya solution
\cite{Bonnor_70}, 
and Hussain solution
\cite{Husain_96}
as special cases. As is apparent from Eq.(\ref{gvmf}), the exterior spacetime cannot be classified in any of the abovementioned subclasses of spacetimes.

Furthermore, satisfaction of the energy conditions puts a constraint on the otherwise free generalized Vaidya mass function $\mathcal{M}(R_c,v)$. For the exterior spacetime to obey the weak energy condition, the function $\mathcal{M}_1(v)$ appearing in Eq.(\ref{gvmf}) should satisfy the following condition:
\begin{equation}
    \mathcal{M}_1(v)\geq \frac{A}{4}\left(\sqrt{13}-3\right)\exp\left(2B+\frac{\sqrt{13}-5}{3}\right).
\end{equation}
Additionally, $\mathcal{M}_1(v)$ should be a monotone decreasing function of the null coordinate $v$. 

Similarly, for the exterior spacetime to obey the dominant energy condition, $\mathcal{M}_1(v)$ is further restricted as follows:
\begin{equation}
    \mathcal{M}_1(v)\geq \frac{A}{4}(\sqrt{61}-6)\exp{\left(2B+\frac{\sqrt{61}-13}{6}\right)}.
\end{equation}
The strong energy condition, however cannot be satisfied throughout the collapse by this spacetime. More insights in the generalized Vaidya spacetime and its energy conditions can be found in the Appendix B.

%The property of such singularity concerning the causal structure and the strength are discussed in the following subsections:

\section{Nature of the ultra high density region}
To check the nature of the strong gravity region formed due to gravitational collapse, as far as its visibility is concerned, we have to investigate the formation of trapped surfaces around the singularity. Trapped surfaces do not form at time $t$ if 
\begin{equation}\label{visibilitycondition}
    \frac{\rho R^2}{3}<1.
\end{equation}
This ensures the positivity of the expansion scalar of the outgoing null geodesic congruence, which in our case is given by
\begin{equation}
    \theta_l=\frac{2}{R}\left(1-\sqrt{\frac{\rho}{3}}R\right).
\end{equation}

We can hence conclude from Eq.(\ref{visibilitycondition}) that for a given $r_c$ (where $r_c$ is the largest comoving radius of the collapsing cloud) when the redefined time $a$ becomes zero, $\theta_{l}>0$ if
\begin{equation}\label{rsquare}
    r_{c}^2<\lim_{a\to 0} \frac{3}{a^2\rho}.
\end{equation}
We see that for the density configuration mentioned in Eq.(\ref{rhoeternal}), the  inequality (\ref{visibilitycondition}) which is the condition to avoid the formation of trapped surface, is satisfied at $a\to 0$. This is because  
\begin{equation}\label{rhoasquarersquare}
    \lim_{a\to 0}\frac{\rho a^2 r^2}{3}=0.
\end{equation}
Hence, there is always a causal connection from any point in the collapsing cloud to an external observer.

\begin{figure}\label{figpd}
\includegraphics[scale=0.35]{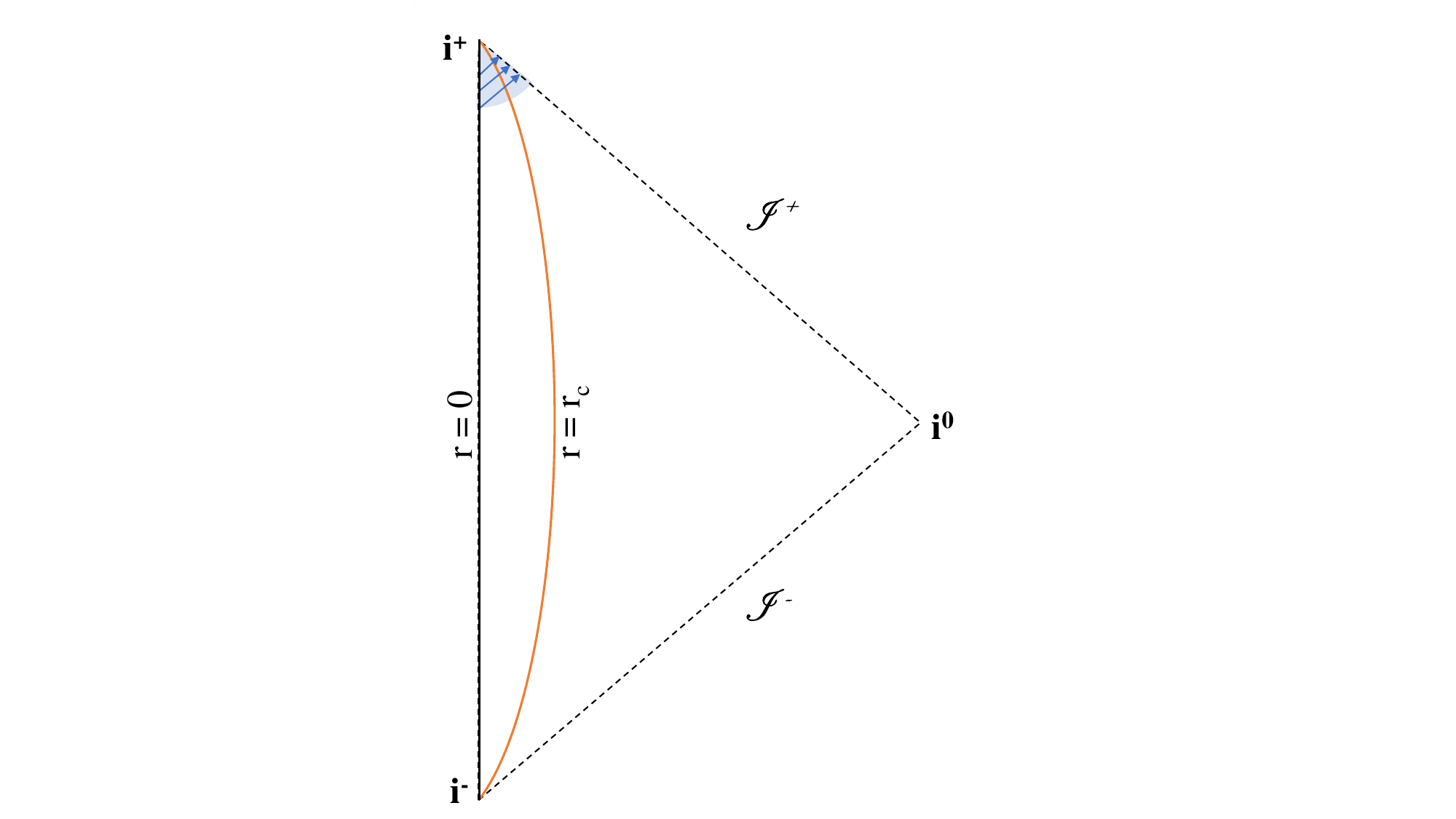}
\caption{Penrose diagram of the eternally collapsing homogeneous scalar field with potential given by Eq.(\ref{V2}) and density given by Eq.(\ref{rhoeternal}). The singularity is formed in future timelike infinity—however, the UHDR forms in a finite comoving time. Since trapped surfaces do not form at all, null geodesics (represented by blue arrows) can escape the UHDR and reach an external observer. This picture holds true if $r_c<\frac{3}{a^2\rho}$ $\forall$ $a\in[0,1]$. Here, $\mathcal{J}^+$, $\mathcal{J}^-$, $i^+$, $i^-$, and $i^0$ are the future null infinity, past null infinity, future timelike infinity, past timelike infinity, and spacelike infinity respectively.}
\end{figure}

As mentioned before, the singularity is formed at the infinite comoving time. One could argue that the knowledge of the causal structure of the singularity is relevant physically (or astrophysically) only if it is formed in a finite comoving time. However, since we have shown that trapped surfaces never form in such eternal collapse, the collapsing region is always visible to the external observer, in principle. The density monotonically increases, and beyond a specific threshold density, the quantum effects dominate. These quantum effects will be hidden behind the event horizon if the collapse ends in a black hole. However, in the scenario discussed here, these effects will be visible since there are no trapped surfaces to trap the outcoming light from the collapsing region. The Penrose diagram 
\cite{Wald_84} 
for such eternal collapse with no formation of trapped surfaces, depicting the causal structure of the UHDR is plotted in Fig.(2). The diagram is for a specific scenario in which there is an upper bound to the comoving radius corresponding to the boundary of the collapsing scalar field. This upper bound is represented by the inequality $r_c<3/\left(a^2\rho \right)$ $\forall$ $a\in [0,1]$. This inequality ensures that $\theta_l>0$ throughout the collapse. The unhindered escape of null geodesics from the UHDR can be traced as seen in the diagram.

Now, at different epochs as we go in the future, along different outgoing radial null geodesics that emanate from the center, we can calculate the quantity $R_{ij} K^i K^j$. It can be shown that as we go forward in time (in other words, as $a$ decreases) $R_{ij}K^iK^j$ increases, blowing up at $a=0$.
\begin{figure}\label{fig2}
\includegraphics[scale=0.5]{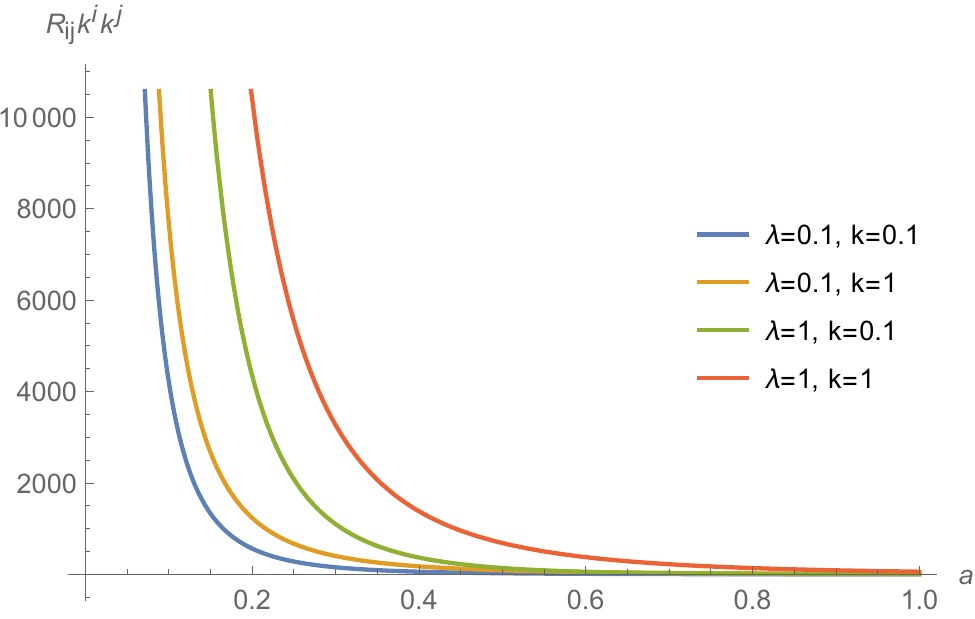}
\caption{$R_{ij}K^iK^j$ along the outgoing radial null geodesic  increases unboundedly as the scale factor decreases in a gravitational collapse of a homogeneous scalar field with potential given by Eq.(\ref{V2}) and density given by Eq.(\ref{rhoeternal}).}
\end{figure}

To see this, consider the geodesic equation of the radial null geodesic escaping from the center written as follows:
\begin{equation}\label{geodesicequation}
    \dot K^{t}+\frac{{K^{r}}'}{a}+\frac{\dot a}{a} K^t=0.
\end{equation}
The outgoing radial null geodesic with the tangent having components
\begin{equation}\label{specialtangents}
  K^t= \frac{1}{a}, \hspace{0.2cm} K^r=\frac{1}{a^2}, \hspace{0.2cm} K^{\theta}=K^{\phi}=0,
\end{equation}
is one of the kind described by Eq.(\ref{geodesicequation}).  For this particular geodesic,
\begin{equation}
    R_{ij}K^i K^j=-\frac{3\dot a}{a^3}+\frac{2\dot a^2}{a^4}+\frac{\ddot a}{a^3}.
\end{equation}
In terms of the scale factor, we can rewrite the above equation using Eq.(\ref{adot}, \ref{addot}, \ref{rhoeternal}) as
\begin{equation}
R_{ij}K^iK^j=\frac{\left(k-\log a \right)}{3a^2}\left(64\lambda \left(-1+3k-3 \log a\right)+24\sqrt{3\lambda}\right).
\end{equation}
One can see that $R_{ij}K^iK^j$ increases progressively as $a$ decreases, and blows up at $a=0$. So whereas the actual singularity is never approached or reached, $R_{ij}K^iK^j$ becomes larger and larger with time, as seen in Fig.(3). What it means is that the projection of the Ricci scalar in the null frame with four-velocity $K^i$ increases monotonically as one progress forward in time. 

\section{Conclusions and Discussions}
%The concluding remarks and open concerns are as follows:

The density of a collapsing scalar field with potential, which is a two-dimensional analog of the ``Mexican hat" shaped Higgs field potential, is not an algebraic function of the scale factor. We hence choose a suitable transcendental function for the density configuration and show that the singularity is formed at an infinite comoving time. However, in a finite but large time, the density of the cloud goes beyond a certain cut-off, above which the laws of quantum gravity governs. The absence of trapped surfaces allows one to observe the UHDR where new physics takes place.  

The collapsing scalar field obeys the strong energy condition only if
\begin{equation*}
    \rho+3p=-\frac{6 \ddot a}{a}>0.
\end{equation*}
Using Eq.(\ref{scalefactor}) and some rearrangement leads to
\begin{equation*}
    \frac{1}{k}>\exp{\left(\frac{8t\sqrt{\lambda}}{3}\right)}
\end{equation*}
for the strong energy condition to hold. Since the collapsing system reaches the UHDR in a finite comoving time (beyond which the laws of physics are not governed by general relativity), the above inequality may or may not hold continually, depending on the values of $k$ and $\lambda$. However, the weak energy condition is always obeyed by the collapsing cloud as seen from Eq.(\ref{rhop}). Furthermore, the collapsing scalar field obeys the dominant energy condition only for positive value of the potential $V$, as seen from Eq.(\ref{rho}) and Eq.(\ref{p}). 

 \begin{figure}\label{fig4}
\includegraphics[scale=0.5]{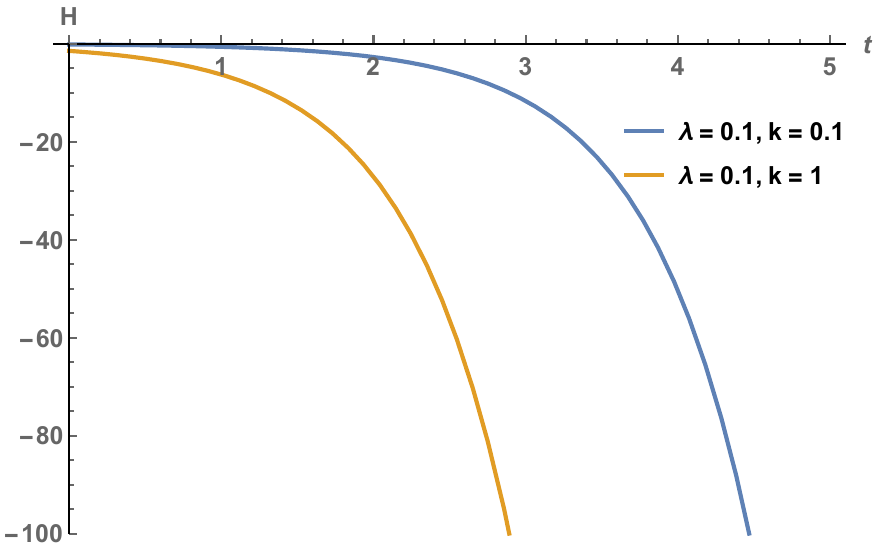}
\caption{The dynamics of $H=\dot a/a$ with time, in a typical gravitational collapse involving a scalar field with potential given by Eq.(\ref{V2}) and density given by Eq.(\ref{rhoeternal})}
\end{figure}
\begin{figure}\label{fig5}
\includegraphics[scale=0.5]{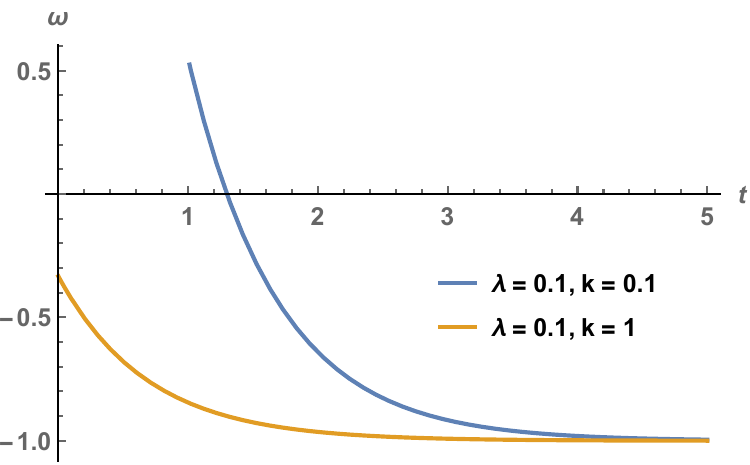}
\caption{The dynamics of $\omega=p/\rho$ with time, in a typical gravitational collapse involving a scalar field with  potential given by Eq.(\ref{V2}) and density given by Eq.(\ref{rhoeternal}).}
\end{figure}

We find that the causal structure of UHDR respects the $\mathbb{Z}_{2}$ (reflection: $\phi \longleftrightarrow -\phi$)  symmetry.  

Unlike massless scalar field collapse, which is proven to have a  non-generic naked singularity
\cite{Christodoulou_94, Christodoulou_99}, we do not know whether the UHDR formed due to the scalar field with a nonzero $\phi^2$ term gives a generic outcome. The visible nature of the UHDR is nevertheless stable under small perturbation in the parameters $\lambda$ and $k$, as seen from Eq.(\ref{rhoeternal}) and Eq.(\ref{rhoasquarersquare}).

It is known that in our case, the Einstein equations and the real scalar field equation both are time-symmetric. Changing $t$ by $-t$ does not change the form of the equations. This fact shows that what we call a gravitational collapse, in the presence of a real scalar field, can yield the time-reversed process of expansion if we retrace the dynamics in a reverse temporal order. If the collapse starts at time $t_i$ and we follow the collapsing process up to time $t_f$ ($t_i > 0$; $t_f > 0$ and $t_f > t_i$) then the expansion in the reverse order starts at $-t_f$ and proceeds up to time $-t_i$. In both cases, time increases monotonically. Expansion in FLRW spacetime in the presence of a scalar field has already been studied in various forms in various models of cosmic inflation 
\cite{Starobinsky_80, Guth_81, Linde_82, Riotto_02}. 
 In inflationary models, spacetime enters the quasi De-Sitter phase after some moments from the initial singularity. For a very brief, finite period, the spatial length scales expand exponentially. During the inflationary phase the ratio $\dot a/a$ remains approximately constant and $p+\rho \sim 0$. Inflation ends when the real scalar field starts to decay into a radiation fluid. One interesting question can be raised at this point: Can the scalar field-driven gravitational collapse in the FLRW spacetime, as presented in this paper, be viewed as the time-reversed form of inflationary expansion? This question becomes meaningful as lots of work has been done on inflation; one may use some of those works in a time-reversed way to predict gravitational collapse. In this regard, we want to point out that the gravitational collapse, in the presence of the potential $V (\phi)$ as given in (\ref{V2}), does not produce inflationary expansion in the time-reversed sense and hence one cannot use models of inflation to predict about the present collapse scenario. In the inflationary models, one has a constant $\dot a /a$ during inflation, whereas Fig.(4) in our case shows $H=\dot a/a$ is a continuously decreasing function of time, for some arbitrary parameter values. If, in reality, the collapse corresponded to time-reversed inflation, $\dot a/a$ should have settled to a constant value in the initial phase of the collapse. The other diagram in Fig.(5) show that in the initial phase of collapse $p/\rho \neq -1$. Consequently, in the time-reversed sense, one cannot identify the time reversed collapse with the inflationary expansion. As a result of these observations, we claim that our model of gravitational collapse, in the presence of a real scalar field having a potential that follows $Z_2$ symmetry, is a new collapse solution that cannot be obtained by time reversing any of the single field inflationary models used till now.
 
 It is worth mentioning that in some models of inflation, called power law inflation \cite{Lucchin_85}, $a(t) \sim t^p$ for some real $p > 1$. In these models, although $\dot a/a$ varies with time but for inflation such models require an exponential potential for the real scalar field. As a result of this we can safely omit these kind of models of inflation as the potential we are working with is not of the exponential type. Nevertheless, the time-reversal of the solution obtained in this paper, still could be interesting as an expanding cosmological model which begins in the infinite past.
%\sout{It seems evident that any nonspacelike geodesic approaching the singularity will only end at the infinite comoving time since the singularity forms at the infinite comoving time. Hence, these geodesics are not incomplete. Apart from its astrophysical implication discussed in the first paragraph of this section,  whether one should consider such so-called singularity, forming at the infinite comoving time, as a severe defect in the manifold depends on how one defines the word ``singularity"
%\cite{Wald_84}. }

It should be noted that we have followed the philosophy of Misner
\cite{Misner_1969}, 
which say that even though we are well aware of the possibility of the failure of general relativity as one approaches the UHDR, one should take into consideration the predictions of general relativity in this regime since it may give us some indication about what one should expect from a more general theory of gravity which works in this regime.

\section{Acknowledgement}
K.M. would like to acknowledge the support of the Council of Scientific and Industrial Research (CSIR, India, Ref: 09/919(0031)/2017-EMR-1) for funding the work.
\begin{widetext}
\section*{Appendix A}
\textbf{Theorem}: In a gravitational collapse involving homogeneous scalar field $\phi(a)$ governed by FLRW spacetime, if $\rho(a)$ is an algebraic function, then either $V(\phi)$ is a transcendental function, or Eq.(\ref{Vorderalpha}) is satisfied. In the latter case, $V(\phi)$ can be algebraic or transcendental.)
    
\textbf{Proof}: If we consider that the density configuration of the collapsing cloud can be expressed as an algebraic function in $a$, then we can say that the density close to the time of formation of the singularity is of the order as follows:
\begin{equation}\label{rhoorder}
    \lim_{a\to 0} \rho \sim O\left(\frac{1}{a^n}\right).
\end{equation}
From Eq.(\ref{adot}), we then obtain
\begin{equation}\label{adotorder}
    \lim_{a\to 0} \dot a \sim O\left(a^{1-\frac{n}{2}}\right),
\end{equation}
differentiating which, we get
\begin{equation}\label{addotorder}
    \lim_{a\to 0} \ddot a \sim O\left(a^{1-n}\right).
\end{equation}
Substituting Eq.(\ref{adotorder}) and Eq.(\ref{addotorder}) in Eq.(\ref{p}), the behavior of the dynamics of pressure close to the time of formation of the singularity is obtained as
\begin{equation}\label{porder}
    \lim_{a\to 0} p \sim O\left(\frac{1}{a^n}\right).
\end{equation}
The above equation along with Eq.(\ref{rhop}) gives
\begin{equation}\label{phidotorder}
 \lim_{a\to 0}   \dot \phi^2 \leq O\left(\frac{1}{a^n}\right).
\end{equation}
Now, since we assume the density to have an algebraic expression in terms of $a$, we imply using Eq.(\ref{porder}) and  Eq.(\ref{rhop}) that $\dot \phi^2$ has an algebraic expression in terms of $a$. We therefore have from Eq.(\ref{phidotorder}) that 
\begin{equation}
     \lim_{a\to 0}   \dot \phi^2\sim O\left(\frac{1}{a^m}\right); \hspace{0.2cm} m\leq n.
\end{equation}
Using the chain rule $\dot \phi^2=\phi_{a}^2\dot a^2$ and Eq.(\ref{adotorder}), we can then conclude
\begin{equation}\label{phiaorder}
    \lim_{a\to 0} \phi_{,a}\sim O\left(\frac{1}{a^{1+\frac{m-n}{2}}}\right).
\end{equation}
Using (\ref{phiaorder}) and (\ref{rhoorder}) in Eq.(\ref{rhofinal}), we obtain the behavior of the potential close to the singularity as
\begin{equation}\label{Vorder}
    \lim_{a\to 0} V \sim O \left(\frac{1}{a^n}\right).
\end{equation}

\textbf{Case I}: If $m=n$, then, 
\begin{equation}\label{phiaorder1}
     \lim_{a\to 0} \phi_{,a}\sim O\left(\frac{1}{a}\right).
\end{equation}
Integrating the expression (\ref{phiaorder1}), we obtain
\begin{equation}\label{phiorder1}
    \lim_{a\to 0} \phi \sim O\left(\log a \right).
\end{equation}
Hence, for (\ref{Vorder}) to hold true, $V$ should be a transcendental function of $\phi$.
%Now consider the potential mentioned in Eq.(\ref{V}). Substituting Eq.(\ref{phiorder1}) in Eq.(\ref{V}), we obtain
%\begin{equation}
 %   \lim_{a\to 0} V \sim O\left(\log a\right)^2 \hspace{1cm}\textrm{or}\hspace{1cm} \lim_{a\to 0} V\sim O\left(\log a\right)^4.
%\end{equation}
%This does not match with the order of $V$ (\ref{Vorder}) we expect when the density %rises as (\ref{rhoorder}).

\textbf{Case II}: If $m<n$, then integrating the expression (\ref{phiaorder}), we obtain
\begin{equation}\label{phiorder2}
    \lim_{a\to 0} \phi \sim O\left(a^{\frac{n-m}{2}} \right).
\end{equation}
Hence, for (\ref{Vorder}) to hold true, (\ref{Vorderalpha}) should be satisfied where
\begin{equation}
    \alpha=\frac{2n}{n-m}.
\end{equation}

\textbf{Corollary}: The density configuration of the collapsing homogeneous scalar field having the potential (\ref{V}) is not an algebraic function of $a$.

This follows because for such potential, substituting either (\ref{phiorder1}) or (\ref{phiorder2}) in (\ref{V}) does not satisfy $(\ref{Vorder})$ which holds true for algebraic $\rho(a)$. The corollary is valid for both $q>0$ and $q<0$ in Eq.(\ref{V}).

\section*{Appendix B}
For a spacetime to be a valid solution of the Einstein's field equations, the stress-energy tensor giving rise to such spacetime, should obey certain energy conditions. Here we investigate the energy conditions for the stress-energy tensor corresponding to generalized Vaidya spacetime as mentioned in Eq.(\ref{gvs}). Consider two null dual vectors $l_{\mu}$ and $n_{\mu}$ in the $(v,R_c,\theta, \phi)$ coordinates such that
\begin{equation}
    l_{\mu}=(1,0,0,0), \hspace{0.2cm} \textrm{and} \hspace{0.2cm} n_{\mu}= \left(\frac{1}{2}\left(1-\frac{2\mathcal{M}}{R_c}\right),1,0,0 \right).
\end{equation}
The stress-energy tensor for generalized Vaidya spacetime can then be expressed as superposition of null dust and perfect fluid as
\cite{Wang_99}
\begin{equation}
    T_{\mu\nu}=T_{\mu\nu}^{(n)}+T_{\mu\nu}^{(m)},
\end{equation}
where
\begin{equation}
    T_{\mu\nu}^{(n)}=\bar{\epsilon}l_{\mu}l_{\nu}, \hspace{0.2cm} \textrm{and} \hspace{0.2cm}   T_{\mu\nu}^{(m)}=(\epsilon+\mathcal{P})\left(l_{\mu}n_{\nu}+l_{\nu}n_{\mu}\right)+\mathcal{P}g_{\mu\nu}.
\end{equation}
Here,  $\bar{\epsilon}$, $\epsilon$, and $\mathcal{P}$ can be expressed in terms of the Vaidya radius $R_c$ and the derivatives of the generalized Vaidya mass function $\mathcal{M}(R_c,v)$ as
\begin{equation}\label{ecparameters}
    \bar{\epsilon}=-\frac{2 \mathcal{M},v}{R_c^2}, \hspace{0.2cm} \epsilon=\frac{2\mathcal{M},_{R_c}}{R_c^2}, \hspace{0.2cm} \textrm{and} \hspace{0.2cm} \mathcal{P}=-\frac{\mathcal{M},_{R_c R_c}}{R_c}.
\end{equation}
The generalized Vaidya solution includes most of the known solutions of the Einstein's field equations
\cite{Wang_99}. For e.g., if one can express the generalized Vaidya mass function as follows:
\begin{equation}\label{seriesgvmf}
    \mathcal{M}(R_c,v)=\sum_{n=-\infty}^{n=+\infty} b_n(v) R_c^n,
\end{equation}
then for $b_i(v)=0$ $\forall$ $i \neq 0$, and $b_0(v) \neq 0$, the well known Vaidya solution
\cite{Vaidya_51} 
is obtained. Now, based on the choice of functions $b(v)$, various other known particular solutions can be achieved like the monopole solution
\cite{Barriola_89}, 
the de-Sitter/ anti de-Sitter solution, charged Vaidya solution
\cite{Bonnor_70}, 
and the Hussain solution
\cite{Husain_96}. It is worth noting that since $\bar{\epsilon}$, $\epsilon$ and $\mathcal{P}$ are linear in terms of derivatives of $\mathcal{M}$, the linear superposition of multiple special solutions is also a solution. Now, since the class of generalized Vaidya mass function as obtained in Eq.(\ref{gvmf}) as a result of smooth matching with the interior solution cannot be expressed in the series form Eq.(\ref{seriesgvmf}), the exterior spacetime in this paper cannot be classified in any of these known solutions.

Now, for the matter field corresponding to generalized Vaidya spacetime to satisfy the weak energy condition, $\bar{\epsilon}\geq 0$ and $\epsilon \geq 0$. These imposes restrictions on $\mathcal{M}(R_c,v)$, such that
\begin{equation}\label{wec1}
    \mathcal{M}_{1,v}\leq 0, \hspace{0.2cm} \textrm{and} \hspace{0.2cm} \mathcal{M},_{R_c}\geq 0.
\end{equation}
Furthermore, one can see from Eq.(\ref{gvmf}) that the above latter inequality restricts the initially free function $\mathcal{M}_1(v)$ as follows:
\begin{equation}
    \mathcal{M}_1(v)\geq E(R_c),
\end{equation}
where
\begin{equation}
    E(R_c)=\frac{A R_c^2}{2}\left(B(2-3B)+2(3B-1)\log (R_c) -3\log(R_c^2)\right).
\end{equation}
Here $B=k+\log r_c$, as mentioned in Eq.(\ref{MAB}). $E(R_c)$ has a maxima at 
\begin{equation*}
    R_c=\exp{\left(B+\frac{\sqrt{13}-5}{6}\right)},
\end{equation*}
and at this value 
\begin{equation}
    E_{\textrm{max}}=\frac{A}{4}\left(\sqrt{13}-3\right)\exp\left(2B+\frac{\sqrt{13}-5}{3}\right).
\end{equation}
Hence, one can get the following restriction on $\mathcal{M}_1(v)$ for the weak energy condition to be satisfied throughout the collapse:
\begin{equation}\label{wec2}
    \mathcal{M}_1(v)\geq E_{\textrm{max}}.
\end{equation}

Strong energy condition demands that, $\mathcal{P}\geq 0$, in addition to the satisfaction of Eq.(\ref{wec1}) and Eq.(\ref{wec2}).This inequality is equivalent to the criteria
\begin{equation}
    \mathcal{M},_{R_c R_c}\leq 0,
\end{equation}
as seen from Eq.(\ref{ecparameters}). However the above inequality cannot be satisfied throughout the collapse, since the maximum value of $\mathcal{M},_{R_c R_c}$, which is obtained at
\begin{equation}
    R_c=\exp{(B-3)},
\end{equation} is 
\begin{equation*}
    13 A \exp{(B-3)},
\end{equation*}
and this is always positive.

For the dominant energy conditon to hold true, $\epsilon \geq \mathcal{P}$. Using Eq.(\ref{ecparameters}), we obtain the inequaltiy
\begin{equation}
    2\mathcal{M},_{R_c}+R_c\mathcal{M},_{R_c R_c}\geq 0.
\end{equation}
This inequality puts a further constraint on  $\mathcal{M}_1(v)$ as follows:
\begin{equation}\label{dec}
    \mathcal{M}_1(v)\geq G_{\textrm{max}},
\end{equation}
where $G_{\textrm{max}}$ is the maximum value of the following function:
\begin{equation}
      G(R_c)=\frac{A R_c^2}{2}\left(7B-1-6B^2+(12B-7)\log (R_c) -6\log(R_c^2)\right),
\end{equation}
$G_{\textrm{max}}$ is obtained at 
\begin{equation*}
    R_c=\exp{\left(B-\frac{13-\sqrt{61}}{12}\right)}.
\end{equation*}
and is given by
\begin{equation}
    G_{\textrm{max}}=\frac{A}{4}(\sqrt{61}-6)\exp{\left(2B+\frac{\sqrt{61}-13}{6}\right)}.
\end{equation}
\end{widetext}
 

\begin{thebibliography}{}

\bibitem{Hawking_73} S. W. Hawking and G. F. R. Ellis, The large scale structure of spacetime, Cambridge University Press (1973).

\bibitem{Wald_84} R. M. Wald, General Relativity. Chicago, USA: Chicago
Univ. Pr., 1984.

\bibitem{Penrose_69} R. Penrose, Riv. Nuovo Cimento Soc. Ital. Fis. \textbf{1}, 252 (1969).

\bibitem{Joshi_07}  P. S. Joshi, \textit{Gravitational Collapse, and Spacetime Singularities}, (Cambridge University Press, Cambridge, England, 2007).

\bibitem{Geroch_70} R. Geroch, Journal of Mathematical Physics, \textbf{11}, 2, 437-449 (1970).

\bibitem{Joshi_82} P. S. Joshi and J. V. Narlikar, Pramana, \textbf{18}, 5 (1982).

\bibitem{Joshi_93} P. S. Joshi and I. H. Dwivedi, Phys. Rev. D \textbf{47}, 5357 (1993). 

\bibitem{Mosani_20(2)} K. Mosani, D. Dey, and P. S. Joshi, Phys. Rev. D  \textbf{102}, 044037 (2020).

\bibitem{Deshingkar_98} S. S. Deshingkar, S. Jhingan, and P. S. Joshi, Gen. Relativ. Gravit. \textbf{30}, 1477 (1998).

\bibitem{Mena_2000} F. C. Mena, R. Tavakol, and P. S. Joshi, Phys. Rev. D, \textbf{62}, 044001 (2000).

\bibitem{Mosani_20} K. Mosani, D. Dey, and P. S. Joshi, Phys. Rev. D \textbf{101}, 044052 (2020).

\bibitem{Christodoulou_91} D. Christodoulou, Comm. Pure Appl. Math. \textbf{44}, 339 (1991).

\bibitem{Magli_97} G. Magli, Class. Quant. Grav. \textbf{14}, 1937 (1997).

\bibitem{Magli_98} G. Magli, Class. Quant. Grav. \textbf{15}, 3215 (1998).

\bibitem{Harada_99} T. Harada, K. Nakao, and H. Iguchi, Class. Quant. Grav. \textbf{16}, 2785 (1999).

\bibitem{Harada_02} T. Harada, H. Iguchi, and K. Nakao, Prog. Theor. Phys. \textbf{107}, 449 (2002).

\bibitem{Goswami_04} R. Goswami and P. S. Joshi, Classical Quantum Gravity \textbf{21}, 3645 (2004).

\bibitem{Goswami_04b} R. Goswami and P. S. Joshi, 	arXiv:gr-qc/0410144.

\bibitem{Giambo_03} R. Giambio, F. Giannoni, G. Magli, and P. Piccione, Commun. Math. Phys. \textbf{235}, 563 (2003).

\bibitem{Giambo_06} R. Giambo, J. Math. Phys. \textbf{47}, 022501 (2006).

\bibitem{Darmios_27} G. Darmois, Les ´equations de la gravitation einsteinienne. Ch. V,M´emorial de Sciences Mathematiques, Fascicule XXV. Paris: Gauthier-Villars (1927).
\bibitem{Israel_67} W. Israel, Nuovo cimento, \textbf{44B}; erratum \textbf{48B} (1967).
\bibitem{Poisson_04} E. Poisson, A Relativist's Toolkit: The Mathematics of Black-Hole Mechanics, (Cambridge University Press, Cambridge, England, 2004).

\bibitem{Christodoulou_94} D. Christodoulou,
Annals of Mathematics Annals of Mathematics, \textbf{140}, 607 (1994).

\bibitem{Christodoulou_99} D. Christodoulou, Annals of Mathematics, \textbf{149}, 183 (1999).

\bibitem{Higgs_64} P. W. Higgs, Phys. Rev. Lett., \textbf{13}, 508 (1964).
Oct 1964.

\bibitem{Melo_17} I. Melo, European Journal of Physics, \textbf{38}, 065404 (2017).

\bibitem{Rugh_02} S.E.Rugh and H. Zinkernagel Studies in History and Philosophy
of Science Part B: Studies in History and Philosophy of
Modern Physics, \textbf{33}, 663 (2002).

\bibitem{Glass_98} E. N. Glass and J. P. Krisch
Phys. Rev. D \textbf{57}, R5945(R) (1998).

\bibitem{Wang_99} A. Wang and Y. Wu, Gen. Relativ. Gravit. \textbf{31}, 107 (1999).
\bibitem{Starobinsky_80} A. Starobinsky, Phys. Letts. B \textbf{91}, 1, 24 (1980).
\bibitem{Guth_81} A. H. Guth, Phys. Rev. D \textbf{23}, 347 (1981).
\bibitem{Linde_82} A. D. Linde, Phys. Letts. B, \textbf{116}, 5, 21 (1982).
\bibitem{Riotto_02} A. Riotto, arXiv:hep-ph/0210162.
\bibitem{Lucchin_85} F. Lucchin and S. Matarrese, Phys. Rev. D \textbf{32}, 1316 (1985).

\bibitem{Misner_1969} C. W. Misner, Phys. Rev. \textbf{186}, 1328 (1969).
 
\bibitem{Vaidya_51} P. C. Vaidya, Proc. Indian Acad. Sci. \textbf{A 33}, 264, Reprint ed  Gen. Relativ. Gravit. \textbf{31} (1999). 

\bibitem{Barriola_89} M. Barriola and A. Vilenkin, Phys. Rev. Lett. \textbf{63}, 341 (1989).

\bibitem{Bonnor_70} W. B. Bonnor and P. C. Vaidya, Gen. Relativ. Gravit. \textbf{1}, 127 (1970).

\bibitem{Husain_96} V. Husain, Phys. Rev. D \textbf{53} , R1759 (1996).

\bibitem{footnote1} Even if there is a discontinuity or jump in the curvature term at $\Sigma$, which violates the junction conditions
\cite{Darmios_27, Israel_67}, 
one could give its physical interpretation in the sense that there
exists surface stress-energy term on $\Sigma$. Such a scenario should not be considered unphysical
\cite{Poisson_04}.
 \end{thebibliography}
\end{document}